%% file: proceedings_NuPhys23.tex
\documentclass[12pt]{article}
\pdfoutput=1
\usepackage[utf8]{inputenc}
\usepackage{amsmath}
\usepackage{amssymb}
\usepackage{graphicx}
\usepackage{slashed}
\usepackage{array}
\usepackage[leftcaption,raggedright]{sidecap}
\usepackage{caption}
\usepackage{subcaption}
\usepackage[hidelinks]{hyperref}
\usepackage{siunitx}
\usepackage[version=4]{mhchem}

\newcommand\pubnumber{NuPhys2023-Cookman}
\newcommand\pubdate{\today}

\newcommand{\beight}{\ce{^{8}B}}

\newcommand{\dmsq}{\ensuremath{\Delta m^{2}_{21}}}
\newcommand{\tonetwo}{\ensuremath{\theta_{12}}}

\def\napoli{King's College London}

\def\support{\footnote{
  Science and Technologies Facilities Council
}}

\def\Title#1{\begin{center} {\Large #1 } \end{center}}
\def\Author#1{\begin{center}{ \sc #1} \end{center}}
\def\Address#1{\begin{center}{ \it #1} \end{center}}

\newcommand\pubblock{\rightline{\begin{tabular}{l} \pubnumber\\
         \pubdate  \end{tabular}}}
\newenvironment{Abstract}{\begin{quotation}  }{\end{quotation}}
\newenvironment{Presented}{\begin{quotation} \begin{center} 
             PRESENTED AT\end{center}\bigskip 
      \begin{center}\begin{large}}{\end{large}\end{center} \end{quotation}}
\def\Acknowledgements{\bigskip  \bigskip \begin{center} \begin{large}
             \bf ACKNOWLEDGEMENTS \end{large}\end{center}}

\input econfmacros.tex

\begin{document}
\begin{titlepage}
\pubblock

\vfill
\Title{Measuring Solar Neutrino Oscillations in the SNO+ Detector}
\vfill
\Author{Daniel Cookman\support}
\Address{\napoli}
\vfill
\begin{Abstract}
  The SNO+ experiment is a large multi-purpose neutrino detector, currently filled with liquid scintillator. For the first time in a single experiment, SNO+ is able to measure the neutrino oscillation parameters \tonetwo{} and \dmsq{} simultaneously through both reactor anti-neutrinos and \beight{} solar neutrinos. The latter approach is demonstrated here, with an analysis of an initial 80 days of scintillator phase data. A Bayesian statistical approach via Markov Chain Monte Carlo is used, allowing for the simultaneous fitting of the oscillation parameters, \beight{} neutrino flux, background components with constraints, and systematic uncertainties. The neutrino oscillation parameter \tonetwo{} was measured to be $38.9^{\circ+8.0^{\circ}}_{\phantom{\circ}-7.9^{\circ}}$, assuming the current global fit flux of \beight{} solar neutrinos. This is consistent with the current global fit result for \tonetwo{}. A sensitivity study shows that this measurement is statistics-limited, and precision could be  improved by a factor of two with two years of livetime, assuming the same backgrounds and selections.

\end{Abstract}
\vfill
\begin{Presented}
NuPhys2023, Prospects in Neutrino Physics\\
King's College, London, UK,\\ December 18--20, 2023
\end{Presented}
\vfill
\end{titlepage}
\def\thefootnote{\fnsymbol{footnote}}
\setcounter{footnote}{0}

\section{Introduction}
SNO+ is a large-scale liquid scintillator detector based \SI{2}{\kilo\metre} underground near Sudbury, Canada. It consists of 780 tonnes of liquid scintillator held within a \SI{6.0}{\metre} radius Acrylic Vessel (AV), surrounding which are 9362 inward-facing PMTs held on a stainless steel PMT Support Structure (PSUP). Between the AV and PSUP is ultra-pure water shielding; the AV is held in place with a series of hold-down and hold-up ropes. For more details about the detector, see~\cite{snopl21}.

\begin{figure}
  \centering
  \includegraphics[width=0.8\textwidth]{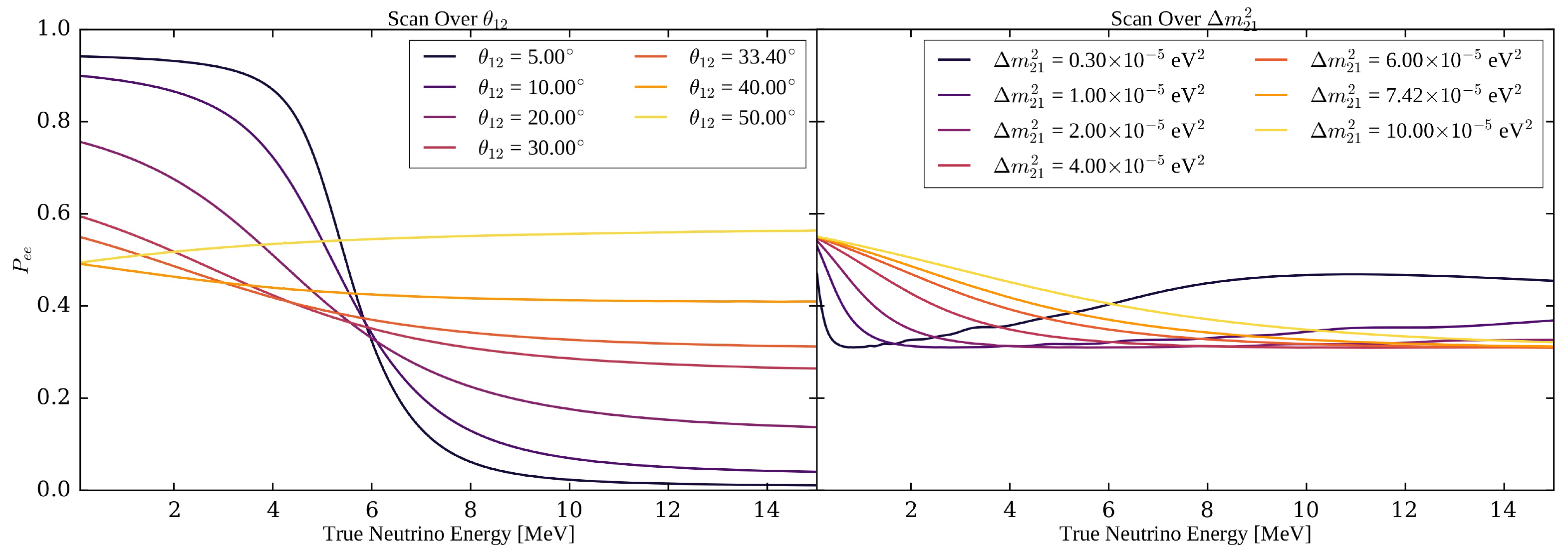}
  \caption{\beight{} solar neutrino survival probability as a function of true neutrino energy, \tonetwo{}, and \dmsq{}. For a given oscillation parameter being scanned over, all other oscillation parameters are set at the NuFit 5.1 global fit values~\cite{esteban20}. Calculations are made using the algorithm described in~\cite{barros11}.}
  \label{fig:th_12_dmsq_21_pee_scan}
\end{figure}

It is possible to observe and measure neutrino oscillations through at least two sources with SNO+: anti-neutrinos from nuclear fission reactors, and neutrinos from the Sun. In both cases, oscillations lead to a deficit in the number of interactions observed compared to no oscillations, with the shape of the measured energy spectrum enabling one to measure the neutrino oscillation parameters \tonetwo{} and \dmsq{}. For the case of solar neutrino oscillations, the survival probability of electron neutrinos ($P_{ee}$) generated within the core of the Sun is impacted by neutrino oscillations in the vacuum of space, as well as the matter effect as the neutrinos pass through the Sun and Earth. This leads to a dependence of $P_{ee}$ on the neutrino's energy, \tonetwo{}, and \dmsq{}, as shown in Fig.~\ref{fig:th_12_dmsq_21_pee_scan}.

This analysis measures \tonetwo{} and \dmsq{} by looking for interactions of \beight{} solar neutrinos in the SNO+ detector via neutrino-electron elastic scattering, $\nu_{x}+e^{-}\to\nu_{x}+e^{-}$. The difference in cross-section between $\nu_{e}$ and $\nu_{\mu,\tau}$, and the (mild) correlation between the energies of the $\nu_{x}$ and scattered electron allow for the measurement of neutrino oscillations through the observed energy spectrum of the scattered electrons.

\section{Method}
Data was selected between 17\textsuperscript{th} May--30\textsuperscript{th} November 2022. A series of cuts was then applied to this data for the purposes of this analysis. After choosing only periods of running for which the detector was operating stably, a number of data cleaning cuts were applied to triggered events to remove instrumental backgrounds. Cosmic ray muon events were also tagged, and a \SI{20}{\second} veto was applied to all subsequent events, to remove cosmogenic backgrounds. To cut out low-energy radioactive backgrounds, triggered events were required to have a reconstructed energy in the range \SIrange{2.5}{14.0}{\MeV}. Backgrounds from sources external to the liquid scintillator, e.g. the AV, external water, and PMTs, were removed through a fiducial volume cut: the reconstructed radial distance of an event from the centre of the AV, $R$, must be less than \SI{5}{\metre}. External backgrounds were further suppresed with the use of two classifiers that distinguish external backgrounds from the solar signal. To substantially reduce the remaining \ce{^{238}U/^{232}Th}-chain and $(\alpha,n)$ backgrounds, a broad coincidence pair tagging algorithm was applied to all events. A PMT hit time classifier was used to search for and remove \ce{^{212/214}BiPo} events for which the two decays occured within one triggered event window.

After all of these cuts, 652 events remained. Considering the application of the \SI{20}{\second} veto, the net livetime of this dataset was 80.615 days. Accompanying this data were full Monte Carlo (MC) simulations of the signal and background processes expected to contribute to the cut data, with the same set of cuts applied to the MC events as in data. Events from data and each MC process after cuts were then binned into 2D histograms in reconstructed energy and radius. The binned MC distributions were then fit to the data using Markov-Chain Monte Carlo (MCMC) via the Metropolis algorithm, with an extended binned maximum likelihood test statistic. This fit allowed the oscillation parameters \tonetwo{} and \dmsq{} to float within the fit, varying the shape of the \beight{} energy distribution. The fit also allowed all signal and background rates, as well as  the reconstructed energy scale systematic to float within constraints. The flux of \beight{} solar neutrinos is constrained using a global fit constraint~\cite{bergstrom16}: $\Phi_{\beight{}} = 5.16^{+0.13}_{-0.09}\times 10^{6}\si{\per\cm\squared\per\second}$.

\section{Results and Future Sensitivity}
\begin{figure}[!th]
  \centering
  \includegraphics[width=0.7\textwidth]{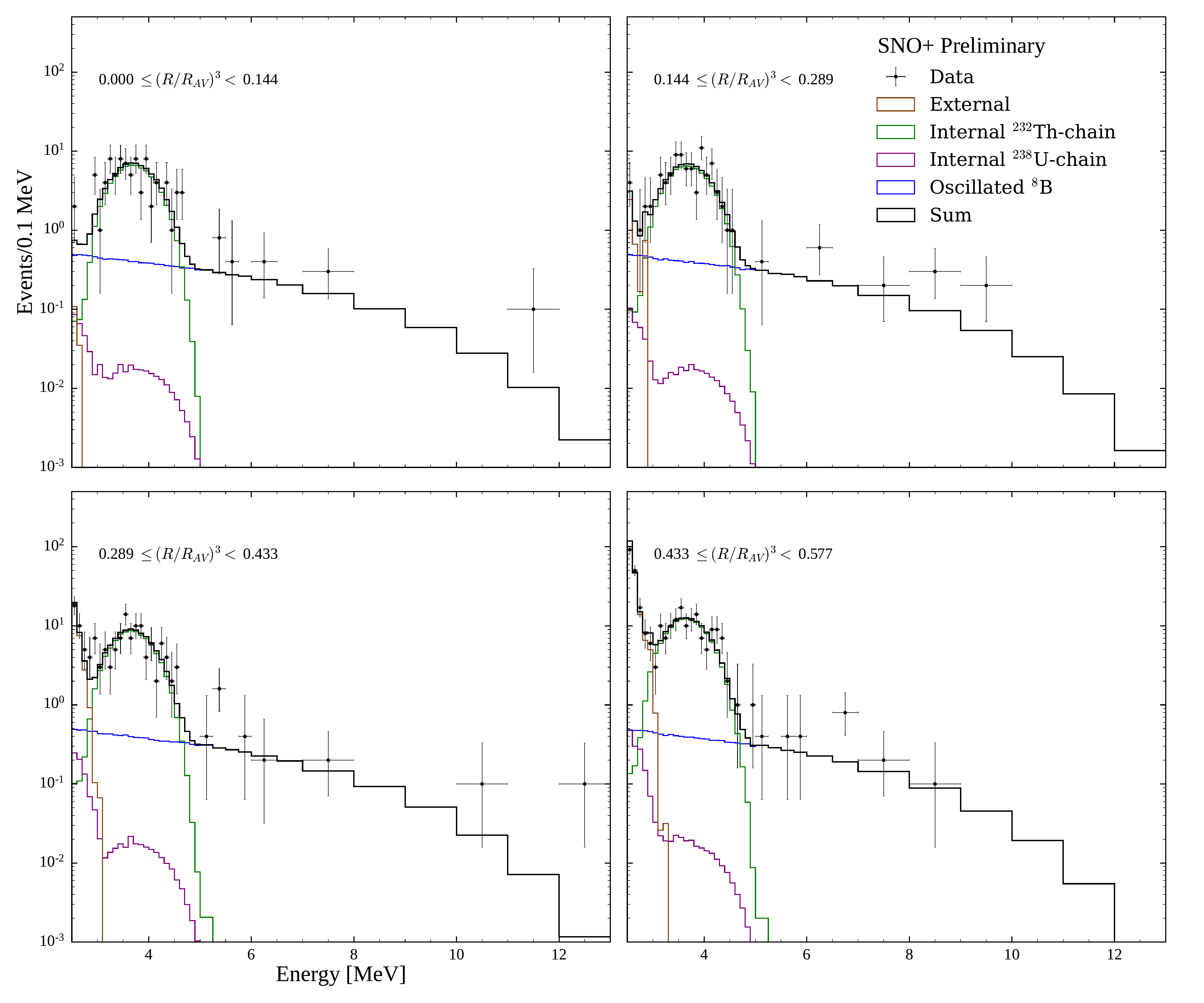}
  \caption{Distribution of selected of events in data in both reconstructed energy and radius, compared to the fitted MC, grouped by type of signal or background. Each subplot corresponds to one equal-volume radial slice.}
  \label{fig:data_mc}
\end{figure}

Fig.~\ref{fig:data_mc} shows the reconstructed energy distributions of data for four equal-volume radial slices, compared to the MC using the fit parameters that obtained the greatest likelihood. The solar signal can be seen to dominate above \SI{5}{\MeV} at all radii, and also around \SI{2.5}{\MeV} in the most central radial bin.

\begin{figure}[!th]
  \centering
    \includegraphics[width=0.7\textwidth]{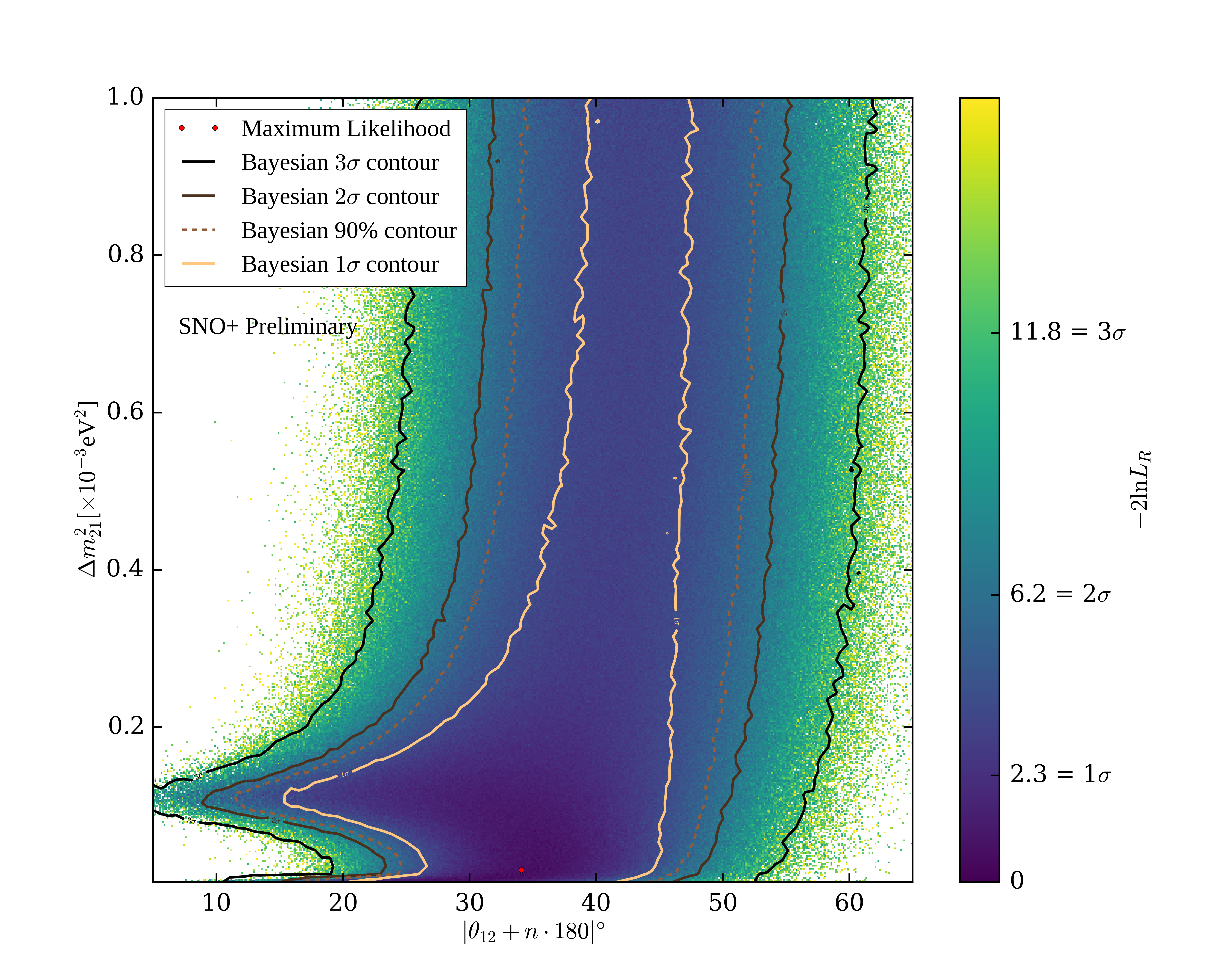}
    \caption{2D marginalised posterior density distribution of \tonetwo{} and \dmsq{} from this analysis. The associated Bayesian CIs are shown, with the colour axis corresponding to twice the negative log-likelihood ratio, $-2\ln{L_{R}}$.}
    \label{fig:osc_param_contours}
\end{figure}

The 2D marginalised posterior density distribution for the parameters \tonetwo{} and \dmsq{} is shown in Fig.~\ref{fig:osc_param_contours}. This plot shows that whilst there is not yet sufficient statistics to constrain \dmsq{}, \tonetwo{} can be measured with the limited dataset. After marginalising onto \tonetwo{}, a Bayesian $1\sigma$ CI of $\tonetwo{} = 38.9^{\circ+8.0^{\circ}}_{\phantom{\circ}-7.9^{\circ}}$ was obtained.

\begin{figure}[!thb]
  \centering
    \includegraphics[width=0.7\textwidth]{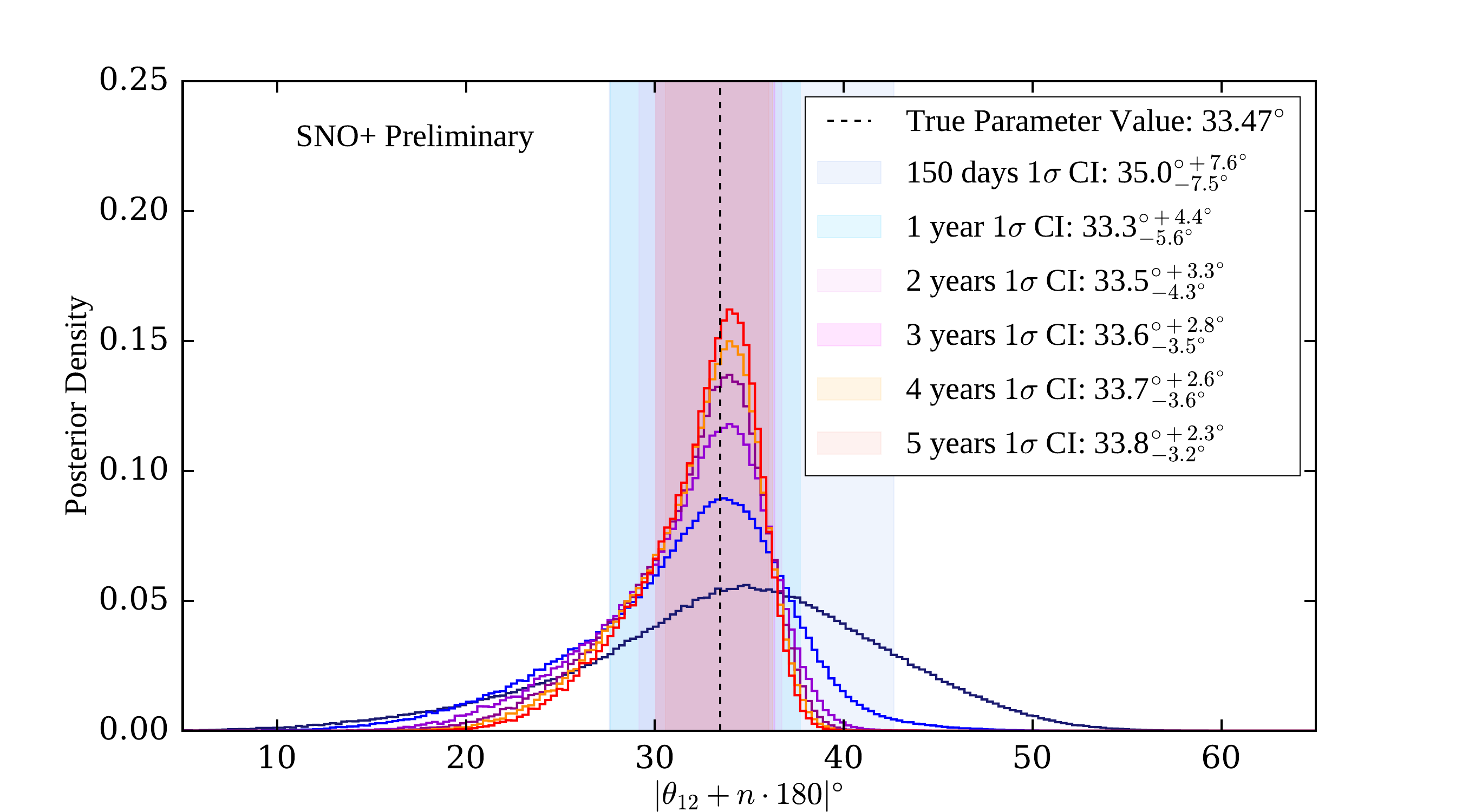}
    \caption{Marginalised posterior density distribution of \tonetwo{} for 6 Asimov datasets of increasing livetime. The Bayesian $1\sigma$ CIs are shown for each livetime scenario.}
    \label{fig:projections}
\end{figure}

A sensitivity study was performed to understand the impact of increased statistics alone on the precision of the oscillation measurement. Asimov fake datasets were generated from MC, for a variety of livetimes between 150 days and 5 years. The rates of each process were determined in general by their fitted rate in the existing dataset. An identical analysis approach was used on the fake datasets as for the real data. The resulting marginalised posterior density distributions for \tonetwo{} are shown in Fig.~\ref{fig:projections}. These show that the measured precision of \tonetwo{} from this analysis can be halved with under two years of livetime, assuming no changes to the detector or analysis.

\section{Conclusions}
An oscillation analysis has been performed with an initial 80 days of data from the SNO+ scintillator phase, by looking at the change in shape of \beight{} solar neutrino-electron elastic scattering events. Using this data, \tonetwo{} was measured as $38.9^{\circ+8.0^{\circ}}_{\phantom{\circ}-7.9^{\circ}}$, consistent with the current global fit value of $33.41^{\circ+0.75^{\circ}}_{\phantom{\circ}-0.72^{\circ}}$~\cite{esteban20}. A sensitivity study indicates that substantial improvements in precision can be made just with an increase in livetime; scintillator-phase data continues to be taken since the dataset in this analysis. More details of this analysis can be found in~\cite{cookman24}. Further increases in precision for this analysis are being explored, such as increasing the fiducial volume, as well as supressing any of the remaining backgrounds.

SNO+ is also measuring \tonetwo{} and \dmsq{} through reactor anti-neutrino oscillations~\cite{morton21}. Because the reactor measurement has greater precision in \dmsq{}, whilst the solar is better in \tonetwo{}, a combined fit between the two approaches could lead to even further improvements in both oscillation parameters compared to each alone.

\Acknowledgements
This work is supported by ASRIP, CIFAR, CFI, DF, DOE, ERC, FCT, FedNor, NSERC, NSF, Ontario MRI, Queen's University, STFC, and UC Berkeley, and have benefited from services provided by EGI, GridPP, and Compute Canada. Daniel Cookman is supported by STFC (Science and Technologies Facilities Council). We thank Vale and SNOLAB for their valuable support.

\providecommand{\href}[2]{#2}\begingroup\raggedright\endgroup
\end{document}

%% file: econfmacros.tex
\def\beq{\begin{equation}}
\def\eeq#1{\label{#1}\end{equation}}
\def\eeqn{\end{equation}}

\def\beqa{\begin{eqnarray}}
\def\eeqa#1{\label{#1}\end{eqnarray}}
\def\eeqan{\end{eqnarray}}

\let\bar=\overbar

\def\Dslash{\not{\hbox{\kern-4pt $D$}}}
\def\dslash{\not{\hbox{\kern-2pt $\del$}}}

\def\msb{{\bar{\ssstyle M \kern -1pt S}}}

